\documentclass[sigconf]{acmart}
\usepackage[htt]{hyphenat}

\renewcommand\footnotetextcopyrightpermission[1]{} 

\usepackage{epsfig}   
\usepackage{epstopdf}
\graphicspath{{./figures/}}

\usepackage{algorithmicx}
\usepackage{algorithm}
\usepackage{amsfonts}
\usepackage[noend]{algpseudocode}

\usepackage{balance}

\newcommand{\algorithmicforeach}{\textbf{foreach}}
\algnewcommand\algorithmicpreset{\textbf{Preset:}}
\algnewcommand\Preset{\item[\algorithmicpreset]}

\algdef{SE}[FOREACH]{ForEach}{EndForEach}[1]{\algorithmicforeach\ #1\ \algorithmicdo}{\algorithmicend\ \algorithmicforeach}%
\algtext*{EndForEach}

\usepackage{textcomp}
\usepackage{listings}
\usepackage{caption}

\begin{document}
    
    
    \title{Developing a Temporal Bibliographic Data Set for Entity Resolution}
    
    \titlenote{This work was partially funded by the Australian Research Council under Discovery Project DP160101934.}

    \author{Yichen Hu}
    \affiliation{Research School of Computer Science, 
        The Australian National University, 
        Canberra ACT 2601, Australia 
    }
    \email{yichen.hu@anu.edu.au}

    \author{Qing Wang}
    \affiliation{Research School of Computer Science,  
        The Australian National University, 
        Canberra ACT 2601, Australia 
    }
    \email{qing.wang@anu.edu.au}
    
    \author{Peter Christen}
    \affiliation{Research School of Computer Science,  
        The Australian National University, 
        Canberra ACT 2601, Australia 
    }
    \email{peter.christen@anu.edu.au}
    
    \begin{abstract}
        Entity resolution is the process of identifying groups of records within or across data sets
        where each group represents a real-world entity. Novel techniques
        that consider temporal features to improve the quality of entity
        resolution have recently attracted significant attention. However,
        there are currently no large data sets available that contain
        both temporal information as well as ground truth information to
        evaluate the quality of temporal entity resolution approaches. In this paper,
        we describe the preparation of a temporal data set based on author
        profiles extracted from the Digital Bibliography and Library Project
        (DBLP). We completed missing links between publications and author profiles in the DBLP data set
        using the DBLP public API. We then used the Microsoft Academic Graph
        (MAG) to link temporal affiliation information for DBLP authors. We
        selected around 80K (1\%) of author profiles that cover 2 million
        (50\%) publications using information in DBLP such as alternative author names and personal web profile to improve the
        reliability of the resulting ground truth, while at the same time
        keeping the data set challenging for temporal entity resolution
        research.  
    \end{abstract}
    
    
    \maketitle
    

    \section{Introduction}
    
    The \emph{Digital Bibliography and Library
        Project} (DBLP) is a computer science bibliography database developed
    and maintained by the University of Trier which has being used by many researchers over the past three decades
    for experimental studies~\cite{ley_dblp_2002}. By
    2018, DBLP contain more than 4.1 million publications\footnote{
        \url{https://dblp.org/statistics/recordsindblp.html}}. While DBLP attempts
    to identify individual authors,
    in many cases errors caused by homonyms or the same names are
    not detected\footnote{~\url{http://dblp.org/faq/}}. 
    DBLP has the potential to be used as a temporal data set with
    publications as records, each of which associates with a time-stamp (i.e. publication date)
    and authors being identified by their unique identifiers. Because authors
    may change their affiliation over time, it is thus challenging to
    identify all records that refer to the same author throughout their
    career. 
    
    Having a large temporal bibliographic data set can be valuable for
    real-time and temporal entity resolution
    research~\cite{altowim_progresser:_2018, ramadan_dynamic_2015}.
    However, so far such a data set has not been made available because
    the public version of DBLP does not contain reliable publication-author
    cross references. Developing such a data set is also challenging due to
    data quality issues around name variations, misspellings, as well as
    frequent names shared by many authors. The version of the DBLP data set used by us is the archive from 1 April, 2018\footnote{\url{http://dblp.org/xml/release/dblp-2018-04-01.xml.gz}}.
    
\begin{figure}[t!]
        \centering
        \includegraphics[width=0.46\textwidth]{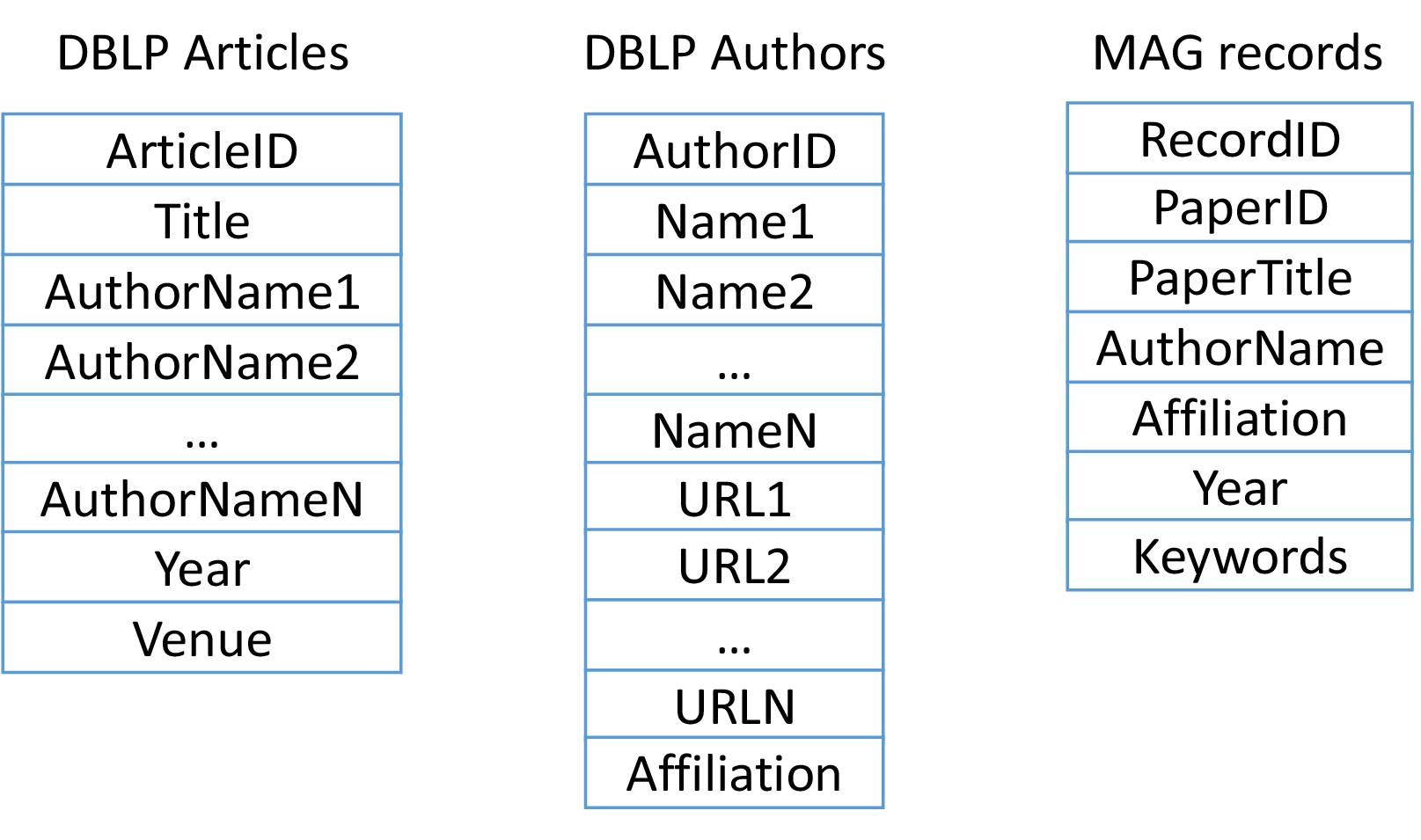}
        \caption{The schemas of the DBLP and MAG source data sets we used
            to create a temporal DBLP data set. Note that there is no
            AuthorID in the DBLP Articles table, and the PaperID in the MAG
            data set is not the same as the ArticleID in the DBLP Articles
            table. Therefore the three data sets are not cross-referred
            with each other.} \label{fig::magdblpcompare}
    \end{figure}

    A second publicly available bibliographic data set, \emph{Microsoft
        Academic Graph} (MAG)~\cite{sinha_overview_2015}, has been released
    for the 2016 KDD cup competition and is also used by us in this paper. The MAG data set used for the KDD cup 2016 and us is a subset that contains records for 2,258,482 publications from the full MAG data set (which contains records for 166,192,182 publications). The MAG data set overlaps with the DBLP data set on a significant proportion, and covers temporal
    affiliation changes which are missing in DBLP. However, unlike DBLP,
    the MAG data set does not contain unique author identifiers.
    Figure~\ref{fig::magdblpcompare} shows the schemas of the DBLP and
    MAG data sets. 
    
    In this paper, we describe our approach to create a temporal data set
    for entity resolution using a subset of relatively
    reliable author profiles from DBLP. We add the temporal
    affiliation information for this refined temporal DBLP data set using
    the MAG data set. Finally, we discuss the characteristics of the proposed
    temporal DBLP data set. To summarize our contributions:
    \begin{itemize}
        \item We create a temporal author entity data set based on DBLP and
        using cross-reference information harvested through DBLP's open
        API. 
        \item We complete a significant proportion of temporal information
        in our proposed temporal data set using affiliation information from
        the MAG data set.
        \item We refine the temporal data set to minimize false matches using
        information in DBLP, including the existence of multiple names of
        the same profile, as well as persistent and non-persistent personal
        URLs. 
        \item We make the generated temporal data set available on \\GitHub\footnote{\url{https://github.com/E-Chen/A-refined-DBLP-temporal-dataset}}.
    \end{itemize}
    
    We next provide necessary background, and in Section~\ref{sec:link}, we discuss the technique we use to create our DBLP temporal data set. In Section~\ref{sec::maglink}, we complete the temporal affiliation information in our temporal DBLP data set, and in Section~\ref{sec::reliability} we address the data quality issues by selecting a subset of author profiles that have either multiple names, static affiliation information, or personal web profiles and are more likely to be accurate. We then describe
    the characteristics of the source and generated data sets in Section~\ref{sec:stat}.
    
    \begin{figure}[t!]
        \begin{lstlisting}[linewidth=\columnwidth,breaklines=true,language=xml, showstringspaces=false]
        <www key="homepages/r/CJvanRijsbergen" ... >
        <author>C. J. van Rijsbergen</author>
        <author>Cornelis Joost van Rijsbergen</author>
        <author>Keith van Rijsbergen</author>
        <note type="affiliation">University of Glasgow, UK</note>
        <title>Home Page</title>
        <url>http://www.dcs.gla.ac.uk/∼keith/</url>
        </www>
        \end{lstlisting}
        \caption{An author record in DBLP.}
        \label{fig::author_example}
    \end{figure} 
    
    
    \section{Background}
    
    DBLP is a popular data set used to evaluate entity resolution
    techniques~\cite{kopcke_evaluation_2010, wang_efficient_2015} as well
    as research in areas such as clustering and bibliometrics.
    However, DBLP is known to have data quality
    issues~\cite{kopcke_object_2014}, such as heterogeneity problems, homonyms, and synonyms.
    
    
    Several recent approaches for temporal entity resolution
    have used various smaller subsets of DBLP. Li et
    al.~\cite{li_linking_2011} used 738 temporal records of 18 authors
    from DBLP, while Wang et al.~\cite{wang_rule-based_2018} used 3,572
    temporal records of more than 20 authors. Chiang et 
    al.~\cite{chiang_tracking_2014} used a larger subset of 100K records
    of an unspecified number of authors, however this data set was only
    used for scalability experiments and not to evaluate entity resolution
    quality. 
    
    The affiliation information in DBLP can be useful to study the
    temporal changes of attribute values in the context of temporal
    entity resolution~\cite{chiang_tracking_2014}. However, it is
    difficult use in a temporal entity resolution scenario because
    the affiliation information in the DBLP is non-temporal, as shown in line 5 in
    Figure~\ref{fig::author_example}. It is therefore currently not possible to
    determine the time period for an affiliation of an author using
    only the DBLP data set. Chiang et al.~\cite{chiang_tracking_2014}
    manually added temporal affiliation information for 258 authors,
    however this approach is expensive and does not scale to the size of DBLP.
    
    
    
    \section{Linking Publications to Unique Author Profiles}
    \label{sec:link}
    
    In this section we discuss how we create temporal
    records for authors in DBLP using the existing DBLP XML file and open
    API\footnote{ \url{http://dblp.org/faq/}}. The desirable form of our temporal data set is a temporally sorted list of publication records
    for each author, where each author has an individual
    record for each of their publications, each author is identified by
    a unique author identifier (ID), and each publication is
    identified by a unique publication ID. 
    
    DBLP is available for public download as an XML file\footnote{
        \url{https://dblp.uni-trier.de/xml/}} which contains two separate
    lists of records, as illustrated in Figure~\ref{fig::magdblpcompare}.
    The first list contains publication records, where each publication
    has a unique \textbf{publication ID} (as \textbf{key} in
    the record), a list of authors described by \textbf{author} tags,
    the \textbf{year} of the publication, and other publication specific
    attributes, such as title, venue etc.~\cite{ley_dblp:_2009}.
    Figure~\ref{fig::ref_example} shows an example publication record
    from the DBLP XML file.
    
    \begin{figure}[t]
        \begin{lstlisting}[linewidth=\columnwidth,breaklines=true,language=xml,showstringspaces=false]
        <inproceedings key="conf/chiir/OBrienFJLTR16" mdate="2016-04-12">
        <author>Heather L. O'Brien</author>
        <author>Nicola Ferro</author>
        <author>Hideo Joho</author>
        <author>Dirk Lewandowski</author>
        <author>Paul Thomas</author>
        <author>Keith van Rijsbergen</author>
        <title>
        System And User Centered Evaluation Approaches in Interactive Information Retrieval (SAUCE 2016).
        </title>
        <pages>337-340</pages>
        <year>2016</year>
        <booktitle>CHIIR</booktitle>
        <ee>http://doi.acm.org/10.1145/2854946.2886106</ee>
        <crossref>conf/chiir/2016</crossref>
        <url>db/conf/chiir/chiir2016.html#OBrienFJLTR16</url>
        </inproceedings>
        \end{lstlisting}
        \caption{A publication record in DBLP.}
        \label{fig::ref_example}
    \end{figure}
    
    The second list contains author records, where each author has a
    unique \textbf{author ID} (as \textbf{key} in the record),
    and \textbf{author} attributes that specify the names that an author
    has used for one of their publication(s). The author identifiers are
    created by the editors of DBLP either manually or using algorithms,
    and have a percentage of unknown errors\footnote{See news on
        2017-09-15 at \url{https://dblp.uni-trier.de/news/}}. Some author
    records, such as the one shown in Figure~\ref{fig::author_example}, have some
    associated URLs recorded (such as a university profile page or a
    Google Scholar profile page, as specified by \textbf{url} tags) and
    are anticipated to be more reliable than those that do not have an
    URL~\cite{ley_dblp:_2009}. Since August 2017, some authors have their
    ORCID imported as well\footnote{
        \url{https://dblp.uni-trier.de/faq/17334571}}. We discuss
    the reliability of author profiles further in Section~\ref{sec::reliability}.
    Some authors have one or more \textbf{affiliation} attributes. However, because no time-stamp is attached to an affiliation, it is not known when an author was a member or worked for a certain
    affiliation.
    
    The DBLP XML file does not contain ID-based cross references between
    the list of publications and the list of authors (as
    Figures~\ref{fig::ref_example} and \ref{fig::author_example}
    show). Author names in the publication list are recorded in plain-text
    and cannot be used to refer to a particular author in the author list
    when there are more than two authors who share exactly the same name.
    The records in the author list have no references to their
    corresponding publication(s) at all. 
    
    To solve this issue, we use the public API provided by
    DBLP to retrieve publication
    records of each author, using author identifiers collected from the
    author list. The API we used in the paper is \\http://dblp.org/pid/\textbf{pid}.xml, where \textbf{pid} refers to an \textbf{author ID}. Each publication record retrieved for each author is in the same
    format as the publication record shown in
    Figure~\ref{fig::ref_example}. However, since the publications are obtained using
    a specific author identifier, we can now link them to a specific
    author profile.
    
    Algorithm~\ref{algo:dblptmprec} shows the procedure used to create a
    temporal data set using DBLP. Since an author can have a number of
    names (as Figure~\ref{fig::author_example} shows), we firstly create
    a $NameIndex$ using the \textbf{AuthorID} of each DBLP author profile
    as index keys, where each key links to its associated author names
    (lines 3 to 6). Note that in line 4
    we used a strategy $IsReliable()$, as discussed in Section~\ref{sec::reliability}, to
    select author profiles that either have multiple names, an affiliation, or have at least one non-persistent or persistent URL.
    
    \begin{algorithm}[t]
        \caption{Create temporal records}
        \label{algo:dblptmprec}
        \begin{algorithmic}[1]
            \Require \hfill\newline
            $\mathbf{E}$ - A list of DBLP author profiles 
            \Ensure \hfill\newline
            $\mathbf{R}$ - A list of DBLP temporal records
            \smallskip
            \State $\mathbf{R} \leftarrow \emptyset $ 
            \State $NameIndex(k, N) \leftarrow emptyindex()$
            \ForEach {$E$ \textbf{in} $\mathbf{E}$}
            \If {$IsReliable(E)$}
            \ForEach {$name$ \textbf{in} $E.names$}
            \State $NameIndex[E.authorid].add(name)$
            \EndForEach
            \EndIf
            \EndForEach
            \ForEach {$k$ \textbf{in} $NameIndex.keys$}
            \State $\mathbf{P} \leftarrow \textit{DBLPAPI}(k)$
            \ForEach {$P$ \textbf{in} $\mathbf{P}$}
            \ForEach {$n$ \textbf{in} $P.authors$}
            \If {$n$ \textbf{in} $NameIndex[k]$}
            \State $R \leftarrow NewRecord(k, n, P)$
            \State $\mathbf{R}.add(R)$
            \EndIf 
            \EndForEach
            \EndForEach
            \EndForEach
            \State $\mathbf{R} \leftarrow SortByDate(\mathbf{R})$ 
            \State \textbf{return} $\mathbf{R}$
        \end{algorithmic}
    \end{algorithm}
    
    \begin{figure}[!t]        
        \begin{lstlisting}[linewidth=\columnwidth,breaklines=true,language=xml, showstringspaces=false]
        <inproceedings key="conf/chiir/OBrienFJLTR16" mdate="2016-04-12">
        <author>Heather L. O'Brien</author>
        <author>Nicola Ferro</author>
        <author>Hideo Joho</author>
        <author>Dirk Lewandowski</author>
        <author>Paul Thomas</author>
        <author>(*\textbf{Keith van Rijsbergen}*)</author>
        <title>
        System And User Centered Evaluation Approaches in Interactive Information Retrieval (SAUCE 2016).
        </title>
        <pages>337-340</pages>
        <year>2016</year>
        <booktitle>CHIIR</booktitle>
        <ee>http://doi.acm.org/10.1145/2854946.2886106</ee>
        <crossref>conf/chiir/2016</crossref>
        <url>db/conf/chiir/chiir2016.html#OBrienFJLTR16</url>
        </inproceedings>
        ...
        <inproceedings key="conf/ictir/ZucconAR11" mdate="2017-05-25">
        <author>Guido Zuccon</author>
        <author>Leif Azzopardi</author>
        <author>C. J. van Rijsbergen</author>
        <title>
        An Analysis of Ranking Principles and Retrieval Strategies.
        </title>
        <pages>151-163</pages>
        <year>2011</year>
        <booktitle>ICTIR</booktitle>
        <ee>https://doi.org/10.1007/978-3-642-23318-0_15</ee>
        <crossref>conf/ictir/2011</crossref>
        <url>db/conf/ictir/ictir2011.html#ZucconAR11</url>
        </inproceedings>
        \end{lstlisting}
        \caption{An example of DBLP API query response using author
            identifier \texttt{homepages/r/CJvanRijsbergen}.}
        \label{fig::api_example}
    \end{figure}

    For each \textbf{AuthorID} $k$ in the $NameIndex$, we obtain a list
    of its associated publications $\mathbf{P}$ using the DBLP API
    (lines 7 and 8). For example, when we
    query the author ID \texttt{homepages/r/CJvanRijsbergen} using the
    DBLP API, we retrieve a list of publication records as shown in
    Figure~\ref{fig::api_example}.
    
    For each author name $n$ from a publication record $P$, we check
    if it exists in the $NameIndex$ associated by $k$. When we can
    find an exact match of a name $n$ in $NameIndex[k]$, we
    create a temporal record $R$ using the author ID $k$, author name
    $n$ of $k$, and the remaining information from publication record
    $P$ (lines 9 to 13).
    
    We create a temporal record for each author of a publication if the author has an author ID in the $NameIndex$, regardless
    of whether the author is the first author or not. We create a Boolean
    attribute \textbf{IsFirstAuthor} to indicate if an author is the
    first author of a certain publication. The reason of creating one
    temporal record for each author is to make the data set more
    interesting for entity resolution, as it will introduce more temporal
    records for each author.
    
    
    
    For example, let an author profile in the DBLP XML file be $A$ =
    \texttt{\{key: homepages/r/CJvanRijsbergen, names: $N$ = [C. J. van
        Rijsbergen, Cornelis Joost van Rijsbergen, Keith van Rijsbergen]\}}.
    We query the author ID \texttt{homepages/r/CJvanRijsbergen} and
    retrieve a list of publication records as
    Figure~\ref{fig::api_example} shows. We compare each name of author
    $A$ against each author name in each publication record. For the
    first reference we can see that \texttt{Keith van Rijsbergen}
    exactly matches a name in $N$, and therefore we then create a
    temporal record: \texttt{\{AuthorID: homepages/r/CJvanRijsbergen,
        PublicationID: conf/chiir/OBrienFJLTR16, AuthorName: Keith van
        Rijsbergen, Year: 2016, CoAuthors: [Heather L. O'Brien,...,Paul
        Thomas], Title: System And User Centered...\}}. We can also see
    that in this example there are six authors, and assuming each of
    them has an author identifier, six temporal records with different
    \textbf{AuthorName}, \textbf{AuthorID} and \textbf{CoAuthors} values
    will be created. 
    
    This approach assumes that there is no publication that has two
    authors with exactly the same name. In other words, while we understand that different authors can share exactly the same name, we assume authors who have exactly the same name are never co-authors of the same paper. For example, if we query for an
    author $A$ who has two names: \texttt{Tom Peter} and
    \texttt{T. Peter}, and obtain a publication record with two
    authors: \texttt{Tom Peter} and \texttt{T. Peter}, we
    will have difficulty to decide which name the author $A$ actually
    used in this publication. When processing the DBLP data set we
    however did not encounter any case where multiple authors shared
    the same name on the same paper.
    
    This DBLP data set is created to simulate a real-world online
    database, where records of individuals are added to the database
    one-by-one in a temporal sequence. Records in our temporal DBLP data
    set are sorted by \textbf{year} and \textbf{month} in ascending
    order. For each publication venue in each year, we currently assign it a
    randomly generated \textbf{month} value. In future work we aim to
    extract the actual publication dates from publication profiles in
    DBLP. 
    
    
    \begin{figure}[t!]
        \centering
        \includegraphics[width=0.46\textwidth]{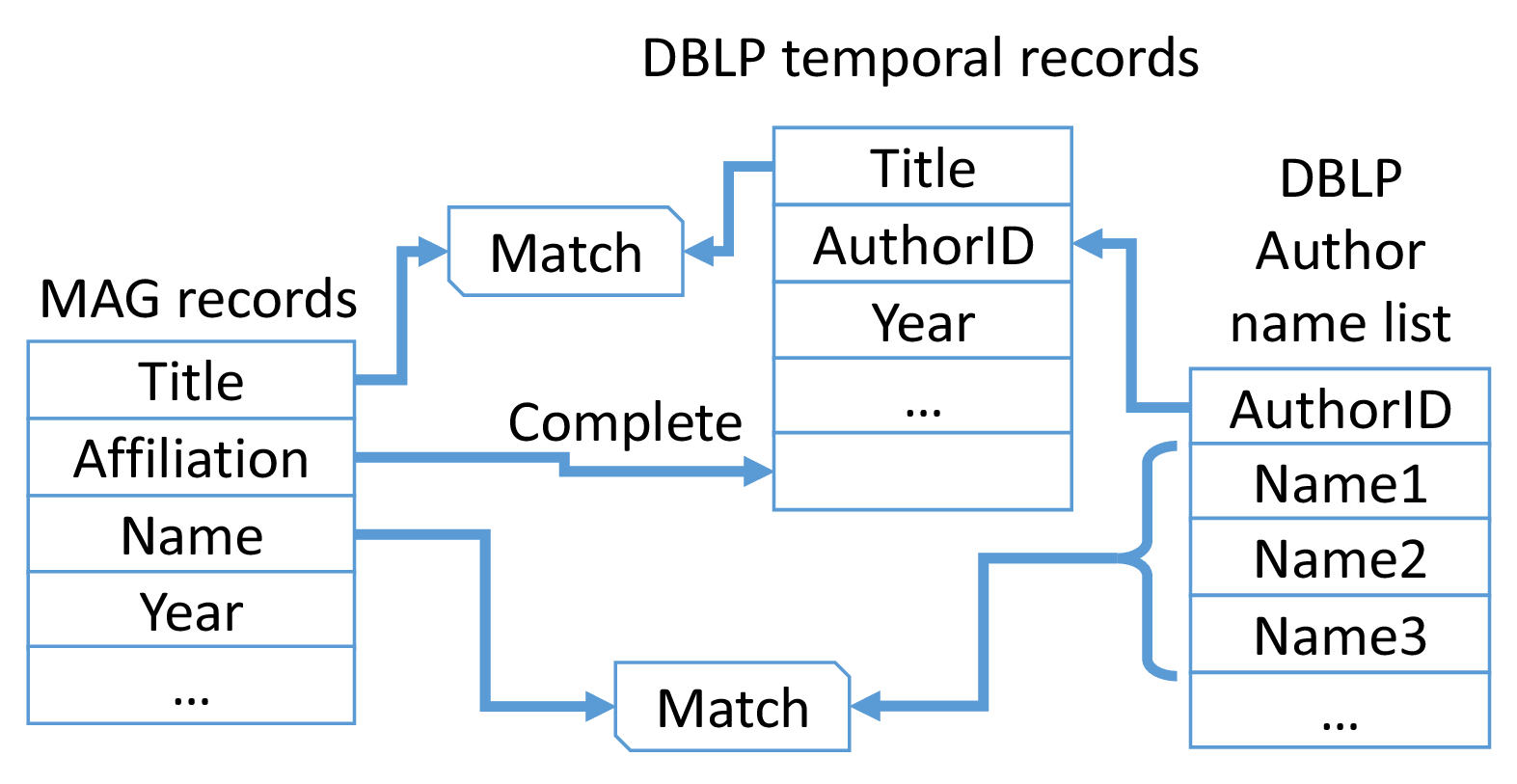}
        \caption{Completing DBLP temporal affiliation information using
            the MAG data set. The article titles in the DBLP temporal records
            are matched to paper titles in MAG records. When a match is
            found, the author name of that MAG record is compared to all
            author names associated to the corresponding DBLP temporal
            record. If a unique matching name pair can be found, the
            affiliation information from the MAG record is added to its
            corresponding DBLP temporal record.}
        \label{fig::mag2dblp}
    \end{figure}
    
    \section{Completing Temporal Affiliation Information using MAG}
    \label{sec::maglink}
    
    From Figure~\ref{fig::author_example}, we can see
    that the affiliation information in DBLP is attached to author
    profiles and does not contain any temporal information. As a result
    we cannot allocate such affiliation information to the DBLP temporal
    records we generated, because we cannot tell when the
    affiliation of an author was valid or had changed. The Microsoft
    Academic Graph (MAG)~\cite{sinha_overview_2015} is an open
    bibliography data set which currently contains 166,192,182
    articles\footnote{\url{https://www.openacademic.ai/oag/}}. The
    records in the MAG data set contain affiliation information at
    different points in time for the authors of each paper. This makes
    it possible to extract temporal affiliation information for authors.
    However, there is no unique identifier for authors in the MAG data
    set, and therefore it is not easy to construct an author-based
    temporal data set using the MAG data set alone. Figure~\ref{fig::mag2dblp}
    shows how we use records in MAG to complete affiliation information
    for DBLP temporal records.

    Algorithm~\ref{algo:affilmag} shows in detail how we complete
    affiliation information for DBLP temporal records, $\mathbf{R}$.
    We first
    map MAG records into an index using their titles
    (lines 1 to 4) and map author IDs to
    their names (lines 5 to 8). Then for
    each DBLP temporal record $R \in \mathbf{R}$, we check if its title
    can be found in the $MAGIndex$ (lines 9
    to 12). When we find a matching title we compare the author name of
    the corresponding MAG record $M$ in the $MAGIndex$ against all names
    that are related to the author ID of $R$. If a matching name can be
    found, we complete the affiliation information of $R$ using the
    affiliation information of $M$. In the $Cleanse()$ function (lines 3
    and 10) we remove all punctuation, spaces, and special characters
    from both DBLP and MAG titles. Using this approach, we were able to match a total of
    418,197 titles across the DBLP and MAG data sets.
    
    \begin{algorithm}[t]
        \caption{Complete DBLP Temporal Affiliation Information Using MAG}
        \label{algo:affilmag}
        \begin{algorithmic}[1]
            \Require \hfill\newline
            $\mathbf{E}$ - A list of DBLP author profiles \newline
            $\mathbf{M}$ - A list of MAG records \newline
            $\mathbf{R}$ - A list of DBLP temporal records
            \Ensure \hfill\newline
            $\mathbf{R}$ - Updated DBLP temporal records
            \smallskip
            \State $\textit{MAGIndex}(title, M) \leftarrow emptyindex()$
            \ForEach {$M$ \textbf{in} $\mathbf{M}$}
            \State $title \leftarrow Cleanse(M.papertitle)$
            \State $\textit{MAGIndex}[title] \leftarrow M$    
            \EndForEach
            \State $NameIndex(k, N) \leftarrow emptyindex()$
            \ForEach {$E$ \textbf{in} $\mathbf{E}$}
            \ForEach {$name$ \textbf{in} $E.names$}
            \State $NameIndex[E.authorid].add(name)$
            \EndForEach
            \EndForEach
            \ForEach {$R$ \textbf{in} $\mathbf{R}$}
            \State $t \leftarrow Cleanse(R.title)$
            \State $id \leftarrow R.authorid$
            \If {$t$ \textbf{in} \textit{MAGIndex}}
            \State $M \leftarrow \textit{MAGIndex}[t]$
            \ForEach {$name$ \textbf{in} $NameIndex[id]$}
            \If {$name$ == $M.name$}
            \State $R.$\textit{affil}$ \leftarrow \textit{MAGIndex}
            [t].$\textit{affil}
            \EndIf
            \EndForEach
            \EndIf
            \EndForEach
            \State \textbf{return} $\mathbf{R}$
        \end{algorithmic}
    \end{algorithm}
    
    
    \section{Selecting Reliable Author Profiles}\label{sec::reliability}
    
    DBLP recently conducted an analysis\footnote{See news on 2017-09-15
        at \url{https://dblp.uni-trier.de/news/}} based on about 70,000
    ORCIDs, which are persistent identifiers for researchers\footnote{For
        details see: \url{http://orcid.org}}. This analysis discovered 600
    records where an author ID is related to more than one ORCID, and
    5,000 records where the same ORCID appears in more than one author profile. These findings suggest that a significant number of author
    IDs are inaccurate in DBLP. If we assume the ORCIDs are accurate
    entity identifiers, then in the case where an author profile $A$ is
    related to more than one ORCID, it indicates that $A$ actually
    contains multiple author profiles which were wrongly linked together.
    When multiple author profiles share the same ORCID, it indicates that
    these author profiles actually refer to the same author and should
    be merged. Note that not all ORCIDs have been incorporated into an
    author profile, and our analysis in Section~\ref{sec:stat} shows that
    only 20,954 ORCIDs have been added to author profiles, and the
    rest of the ORCIDs are still pending validation.
    
    Since we expect the combined data set to be used to evaluate entity
    resolution techniques, it is important to reduce the number of false
    matches and false non-matches (missing true matches), where false matches and missing true matches will reduce both precision and recall~\cite{Chr12,Han17}. By examining
    the original DBLP data set, we discovered several types of
    information that can be used to refine a more reliable subset of
    author profiles. 
    
    \setlength{\abovecaptionskip}{0.5pt}
    \begin{figure}[t]
        \begin{lstlisting}[linewidth=\columnwidth,breaklines=true,language=xml, showstringspaces=false]
        <www key="homepages/189/7254" ... >
        <author>Jeffrey A. McDougall</author>
        </www>
        \end{lstlisting}
        \caption*{(a) The most common type of profile in DBLP.}
        \medskip
        \begin{lstlisting}[linewidth=\columnwidth,breaklines=true,language=xml, showstringspaces=false]
        <www key="homepages/77/10481" ... >
        <author>Emitza Guzman</author>
        <author>Adriana Emitzá Guzmán Ortega</author>
        <note type="affiliation">Technical University Munich, Germany</note>
        </www>
        \end{lstlisting}
        \caption*{(b) A profile with multiple author names.}
        \medskip
        \begin{lstlisting}[linewidth=\columnwidth,breaklines=true,language=xml, showstringspaces=false]
        <www key="homepages/c/PeterChristen" ... >
        <author>Peter Christen</author>
        <url>http://cs.anu.edu.au/~Peter.Christen/</url>
        <note type="affiliation">The Australian National University</note>
        </www>
        \end{lstlisting}
        \caption*{(c) A profile with a non-persistent URL.}
        \medskip
        \begin{lstlisting}[linewidth=\columnwidth,breaklines=true,language=xml, showstringspaces=false]
        <www key="homepages/99/3847-1" ... >
        <author>Wei Song</author>
        <note type="affiliation">
        University of New South Wales, School of Computer Science and Engineering, Sydney, Australia
        </note>
        <url>https://orcid.org/0000-0001-7573-3557</url>
        </www>
        \end{lstlisting}
        \caption*{(d) A profile with a persistent URL.}
        \bigskip
        \caption{Four categories of author profiles in DBLP.}
        \label{fig::4evidences}
    \end{figure}  
    
    We identified four categories of author profiles in the DBLP data set
    with different levels of reliability:
    \begin{enumerate}
        \item \textbf{No support}: Figure~\ref{fig::4evidences} (a) shows
        the most common type of profile which has only one author ID and
        one author name. More than 99\% (1,984,904) of author profiles in
        DBLP are of this type. They are likely created for either
        temporary and non-persistent researchers or they are missed true matches to
        an existing profile. We consider this type of profiles to have no
        support and to be the least reliable. 
        \item \textbf{Multiple names or affiliation}: 38,035 profiles have
        more than one author name or affiliation information as
        Figure~\ref{fig::4evidences} (b) shows, suggesting these profiles
        have been merged by the DBLP team either manually or automatically
        using an algorithm. Name changes and variations are important
        aspects that make a data set realistic for evaluating temporal
        entity resolution techniques, and profiles with multiple names
        have considerable benefit to be included into our temporal data set.
        \item \textbf{Non-persistent URLs}: 27,146 profiles contain at least
        one non-persistent URL, such as a staff profile page from a
        university, as Figure~\ref{fig::4evidences} (c) shows.
        Non-persistent URLs are not meant to be used as author identifiers,
        but they can be used as relatively strong evidence that a profile
        received fair attention and scrutiny.
        \item \textbf{Persistent URLs}: 27,496 profiles contain verified
        persistent identifiers from a third party, such as Google Scholar,
        ORCID, or Scopus\footnote{ \url{https://www.scopus.com/}} as
        Figure~\ref{fig::4evidences} (d) shows. We consider profiles with
        persistent URLs to be the most reliable.
    \end{enumerate}
    
    In the temporal data set we aim to develop, we create temporal records using
    author profiles that either have multiple names, an affiliation, or
    have at least one non-persistent or persistent URL (i.e.\ these
    profiles are in one of the last three categories). We call these criteria \textit{support information}.
    
    
    \section{Characteristics of Data sets}
    \label{sec:stat}
    
    \begin{figure}[t]
        \centering
        \includegraphics[width=0.46\textwidth]
        {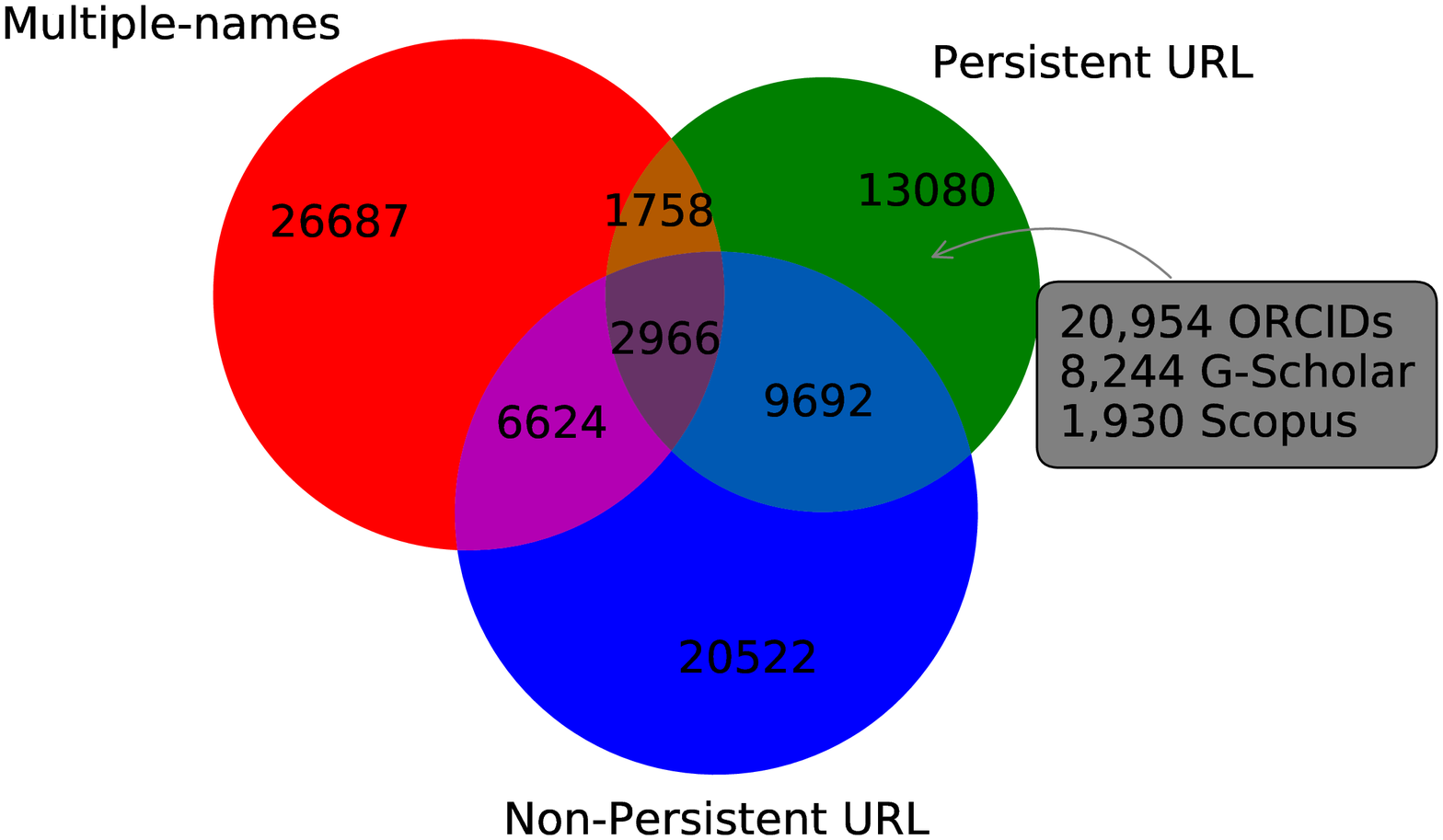}
        \caption{Distribution of support information in author profiles
            in the proposed DBLP temporal data set.}
        \label{fig::info-distribution}
    \end{figure}
    
    In this section we discuss some of the statistics and characters of the
    generated temporal data set, as well as the MAG and the DBLP data
    sets in general.

    Figure~\ref{fig::info-distribution} shows the distribution of author
    profiles that have at least one of the three categories of support
    information described in the previous section. We can see that we
    have a majority of author profiles supported by at least one URL,
    and more than 15K profiles have at least two types of support
    information. Also note that though the DBLP team imported about
    70K ORCIDs into DBLP, only about 20K of them can be found in DBLP
    author profiles, where the remainder of ORCIDs are located in
    publication records and are waiting to be linked to an author
    profile.
    
    
    
\begin{figure}[t]
        \centering
        \includegraphics[width=0.49\textwidth]{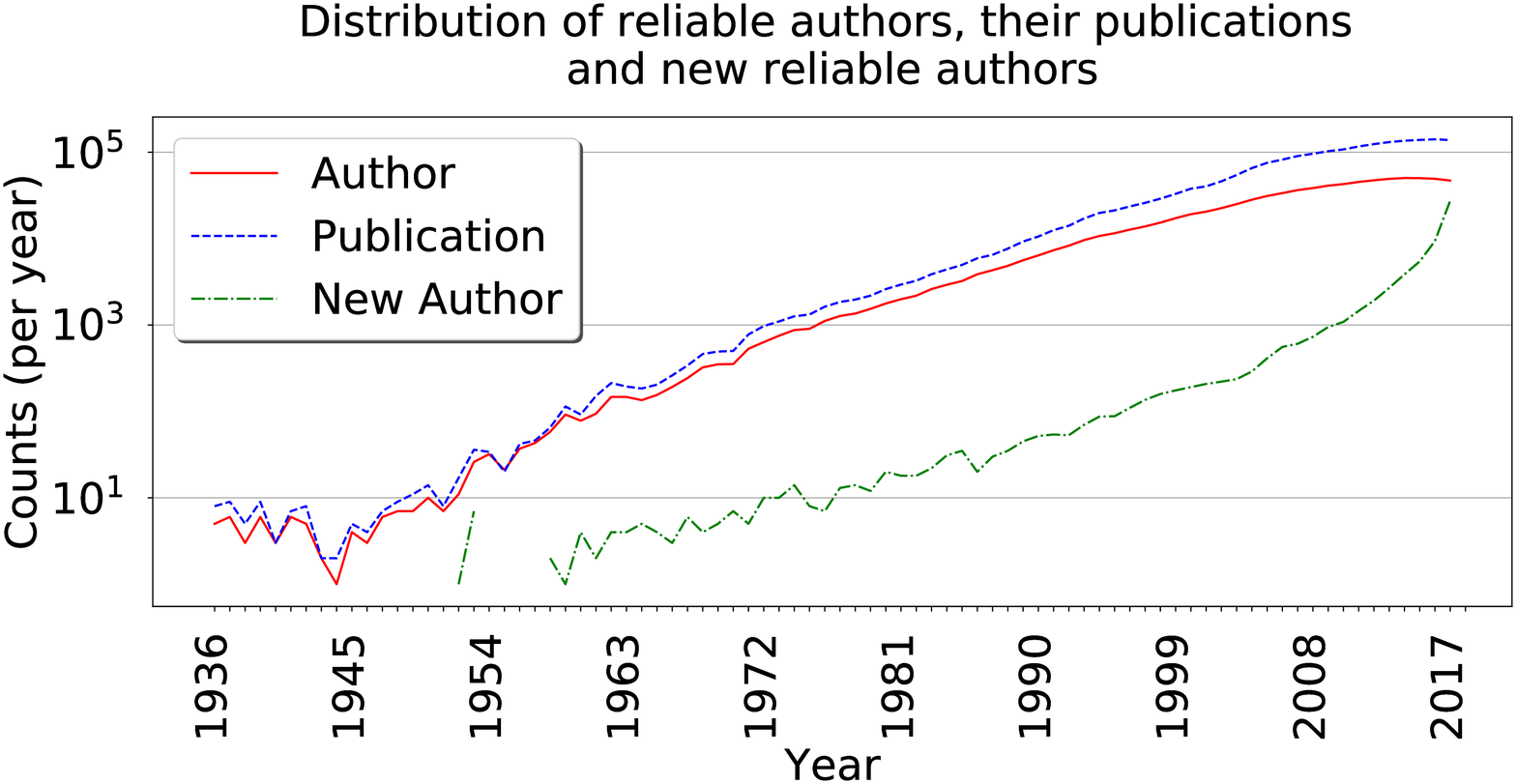}
        \caption{Number of author profiles, publications, and new author
            profiles in each year in the generated temporal DBLP data set.}
        \label{fig::3linechart}
    \end{figure}

    Figure~\ref{fig::3linechart} shows the number of author profiles,
    publications, and new author profiles by year. A new author profile
    refers to a profile that was newly added to a data set. The number of new authors increased sharply in the
    last ten years while the total number of authors and publications
    increased steadily over time. 
    
    
    
\begin{figure}[t]
        \centering
        \includegraphics[width=0.5\textwidth]{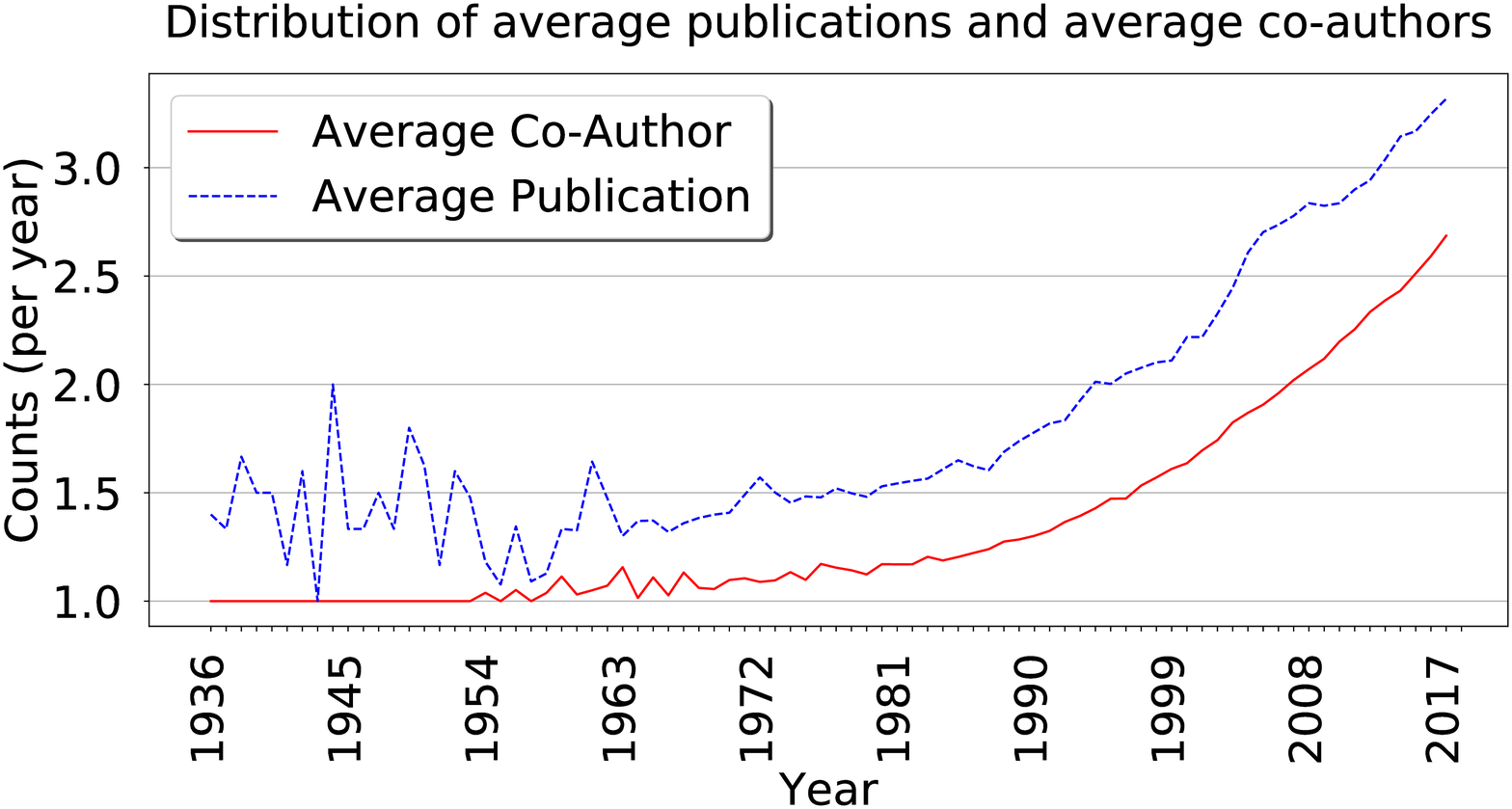}
        \caption{Average number of publications per author and average
            co-authors per publication in each year in the generated
            temporal DBLP data set.}
        \label{fig::2lineyearchart}
    \end{figure}

    Figure~\ref{fig::2lineyearchart} shows the average number of publications for each author per year and the average number of co-authors for each publication each year. Both of these have grown steadily since 1990. Authors are collaborating more over time suggesting that research networks and communities are getting more complex over time. 
    
\begin{figure}[t]
        \centering
        \includegraphics[width=0.5\textwidth]{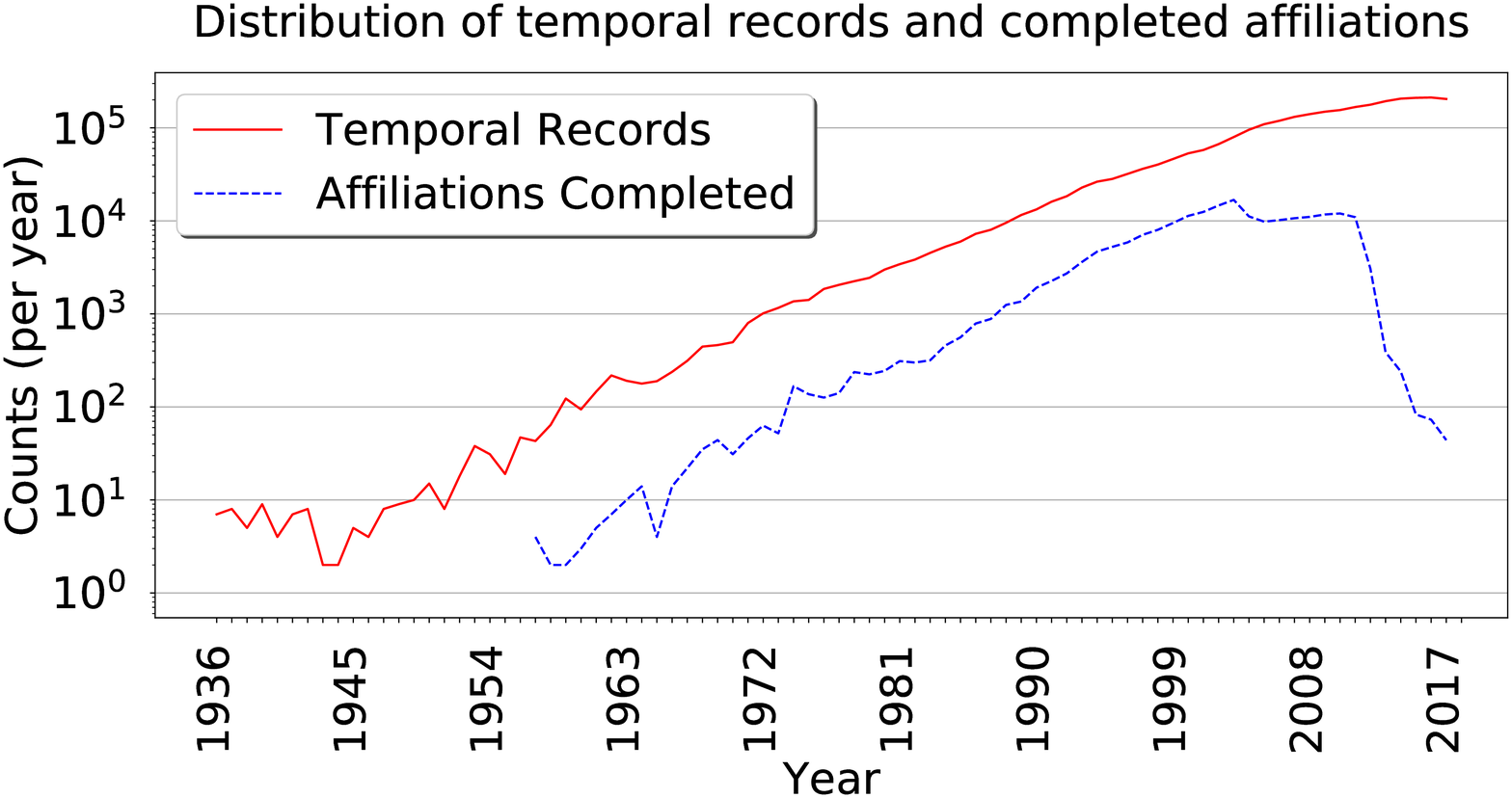}
        \caption{Number of temporal records and records with affiliation
            information completed using MAG in each year in the generated
            temporal DBLP data set.}
        \label{fig::2linerecordchart}
    \end{figure}

    Figure~\ref{fig::2linerecordchart} shows the number of temporal records that have their affiliation information completed in our generated data set. We can see that a large percentage of temporal records do not have their affiliation information completed. As mentioned in Section~\ref{sec::maglink}, we managed to match 418,197 titles from which we completed affiliation information for 205,490 temporal records. Note that the MAG data set we used was a snapshot provided in 2013, and therefore papers published after 2013 cannot be linked. In the future we plan to use an updated and completed version of the MAG data set\footnote{ \url{https://www.openacademic.ai/oag/}} to complete the affiliation information of more authors.
    
    \begin{table*}[t]
        \centering
        \caption{Statistics of MAG data set}
        \label{table::magstat}
        \begin{tabular}{|l|c|l|c|c|c|}
            \hline
            & Total &                 & Avg & Median & Max \\ \hline
            Number of unique name strings        &     2,316,982  	& Per publication      &  5.56   &  3   &  3203   \\ \hline
            Number of publications              &     2,258,482      & Per unique name string      &   5.42  &  1 & 3360  \\ \hline
            Number of unique affiliation strings		&    771,997 & Per unique name string      &  1.77  &   1  &   565  \\ \hline
        \end{tabular}
        
        \bigskip
        
        \centering
        \caption{Statistics of the DBLP XML data set}
        \label{table::dblpstat}
        \begin{tabular}{|l|c|l|c|c|c|}
            \hline
            & Total &                 & Avg & Median & Max \\ \hline
            Number of authors                   &  2,066,233    & Per unique name string &  1.01   &  1   &   141  \\ \hline
            Number of unique name strings        &     2,082,526  	& Per author      &  1.02   &  1   &  10  \\ \hline
            Number of unique name strings  		&  	-	& Per publication &   2.93  &  3   &  287   \\ \hline
            Number of publications              &  4,011,876   & Per unique name string      &   5.72  & 2   &  3133  \\ \hline
        \end{tabular}
        
        \bigskip
        
        \centering
        \caption{Statistics of the generated DBLP temporal data set}
        \label{table::tempstat}
        \begin{tabular}{|l|c|l|c|c|c|}
            \hline
            & Total &                 & Avg & Median & Max \\ \hline
            Number of authors                   &  81,177    & Per unique name string &  1.05   &  1   &   69  \\ \hline
            Number of authors         					&  	-	& Per publication &  1   &  1.5   &   37  \\ \hline
            Number of authors 					&   -	& Per venue  &   112.6   &  26   &  19,703   \\ \hline
            Number of authors				&   -	& Per unique affiliation string &  2.47   &  1   &   167  \\ \hline
            Number of unique name string        &    112,166   	& Per author      &  1.43   &  1   &  10   \\ \hline
            Number of publications              &  1,956,963   & Per author      &  36.1   & 16  &  1,273  \\ \hline
            Number of unique affiliation string		&    88,773  		& Per author      &   2.7   &  1   &   136     \\ \hline
        \end{tabular}
    \end{table*}
    
    Table~\ref{table::tempstat} shows the statistic of the generated 
    temporal DBLP data set. In total we generated 2,931,038 temporal records for
    81,177 authors and 1,956,963 publications between 1936 to 2018. In contrast to the temporal DBLP data set
    without filtering unreliable profiles (Table~\ref{table::dblpstat}),
    the proposed data set contains only 1\% of profiles while it covers
    about 50\% of all publications in DBLP, suggesting that authors who have a relatively more developed DBLP profile are more likely to be regular authors. Since the MAG data set does
    not have unique author identifier, statistics for author profiles
    are not available for the MAG data set.
    
    The most commonly shared author name is \texttt{Wei Wang}, which is shared by 69 different authors as Table~\ref{table::tempstat} shows. The author with the largest number of names is Naufal M Saad, who has 10 different names:
    \texttt{Muhammad Naufal Bin Muhammad Saad, Naufal M Saad,..., Mohammed Naufal bin Mohamad Saad, M Naufal Mohamad Saad}. The venue with the largest number of authors is CoRR, which involves 19,703 authors. 18,047 authors do not have a publication as the first author.
    
    Many authors have multiple different affiliation strings in MAG, but when we
    inspect these affiliations, most of them are different names of the
    same institution, for example, \texttt{Computer Vision LabETH
        Zurich DITET  BIWI}, \texttt{Computer Vision Laboratory ETH
        Zurich}, and \texttt{ETH Zurich Computer Vision Laboratory
        Sternwartstrasse 7 8092 Switzerland} are some affiliation strings
    for the same author. This suggests that a user of our data set
    may wish to standardize these affiliation strings before using the
    data set. To avoid variance caused by different standardization
    techniques and to keep the quality and variability of the data set,
    we did not apply standardization to the generated data set.
    
    
    \section{Conclusion and Future Work}
    
    We have described the development of a temporal data set for entity resolution which
    was created using the publicly available \emph{Digital Bibliography
        and Library Project} (DBLP) and the \emph{Microsoft Academic Graph}
    (MAG) data sets. We used DBLP's public API to link author profiles
    and publication records, and then created one temporal record for
    each author of each publication in DBLP. We then matched titles
    from MAG records to this temporal DBLP data set to complete the
    affiliation information for each author for each publication
    record. We used three categories of support information to refine the
    DBLP author profiles to improve the quality of the proposed temporal
    data set. We generated a temporal data set with 2,931,038 records for
    81,177 authors and 1,956,963 publications between 1936 to 2018, where
    most of the temporal records refer to publications after 1990. The
    data set is made freely available on GitHub for public
    use\footnote{\url{https://github.com/E-Chen/A-refined-DBLP-temporal-dataset}}.
    
    So far we have used only a subset of the large MAG which was made
    available for the KDD competition in 2016. A larger and more complete
    version of MAG~\cite{sinha_overview_2015} which has been linked to AMiner~\cite{tang_arnetminer:_2008} has recently been made available. We aim
    to create a larger and more comprehensive temporal data set by
    linking DBLP to the full MAG and AMiner data sets in the future. 
    
    We also plan to use similarity comparison algorithms, such as
    Jaccard similarity or edit distance~\cite{Chr12}, to match titles
    between MAG and DBLP. We noticed some data parsing issues we had
    with the DBLP XML file that may result in some multi-line titles being
    ignored, and we aim to fix this issue in the future.
    
    \balance
    
    \bibliographystyle{splncs03}    
    \bibliography{My_paper}
    
\end{document}